\documentclass[10pt,letterpaper]{article}
\usepackage{opex3}
\usepackage{color}
\usepackage{float}
\usepackage{afterpage}              
\usepackage{amssymb}
\usepackage{amsmath}

\begin{document}

\title{The effect of input phase modulation to a phase-sensitive optical amplifier}

\author{Tian Li,$^{1,*}$ Brian E. Anderson,$^{1}$ Travis Horrom,$^1$ Kevin M. Jones,$^2$ and Paul D. Lett$^{1,3}$}

\address{$^1$Joint Quantum Institute, National Institute of Standards and Technology and the University of Maryland, College Park, MD 20742 USA\\
$^2$Department of Physics, Williams College, Williamstown, Massachusetts 01267 USA\\
$^3$Quantum Measurement Division, National Institute of Standards and Technology, Gaithersburg, MD 20899 USA}

\email{$^*$litian@umd.edu} 



\begin{abstract*}
Many optical applications depend on amplitude modulating optical beams using devices such as acousto-optical modulators (AOMs) or optical choppers. Methods to add amplitude modulation (AM) often inadvertently impart phase modulation (PM) onto the light as well. While this PM is of no consequence to many phase-insensitive applications, phase-sensitive processes can be affected. Here we study the effects of input phase and amplitude modulation on the output of a quantum-noise limited phase-sensitive optical amplifier (PSA) realized in hot $^{85}$Rb vapor. We investigate the dependence of PM on AOM alignment and demonstrate a novel approach to quantifying PM by using the PSA as a diagnostic tool. We then use this method to measure the alignment-dependent PM of an optical chopper which arises due to diffraction effects as the chopper blade passes through the optical beam.
\end{abstract*}



\section{Introduction}
%
Acousto-optical modulators (AOMs) and electro-optic modulators (EOMs) are standard devices used in optics laboratories for frequency-shifting, amplitude-modulating, and phase modulating optical fields~\cite{PhotonicDevices}. EOMs can provide phase modulation (PM) using the electro-optic response of a crystal, and can provide amplitude modulation (AM) when combined with polarizers. Achieving pure PM or AM using EOMs requires extreme care. Due to effects such as frequency-dependent interference and polarization rotation in the birefringent crystals, PM is often accompanied by residual AM and vice versa~\cite{Whittaker:85, Cusack:04}. Techniques have been developed for combining multiple EOMs to impart an arbitrary mixture of AM and PM on light or to suppress the unwanted modulation~\cite{Cusack:04}.

Driving an AOM with modulated radio frequency can also be used to add AM to the output light in either the zeroth or first diffracted order. This method can also introduce some amount of PM to the light due to changes to the index of refraction in the AOM crystal, such that the optical phase follows the acoustic phase~\cite{HENDERSON1970223, Li_OptLett2005}. This situation, however, is rarely discussed. Many experiments and applications using AOMs are either phase-insensitive or otherwise unaffected by residual PM. Nevertheless, certain phase-sensitive processes are affected. Phase modulation from an AOM has been shown to be an experimental difficulty in some optical phase-sensitive amplifier (PSA) experiments~\cite{Corzo:11, CorzoPRL:12}. In particular, while the PSA can perform completely noiseless amplification of a particular field quadrature, it can also convert PM to AM, making signal-to-noise ratio (SNR) measurements hard to interpret.  It can even lead to apparent increases in the SNR after amplification if inadvertent PM is closely tied to an applied AM signal, as is the case in using many modulation devices.

While AM can be measured with direct detection methods, PM can only be detected using more complicated phase-sensitive or interferometric measurement techniques, and can therefore be difficult to detect and eliminate. Common techniques for measuring the AM and PM of optical beams include homodyne and heterodyne detection~\cite{Cusack:04,bachor_guide_2004}. More indirect methods also exist for converting PM to AM such as using differential absorption in a sample~\cite{Whittaker:85}, reflected light from a cavity~\cite{Yam:15}, phase conjugation methods~\cite{PICHE1988146}, or Brillouin scattering~\cite{Yao98}. Phase modulation is often used in optical communication, such as in phase-shift keying. Signals from phase-shift keying are often demodulated using homodyne or heterodyne techniques and phase sensitive amplifiers have been investigated for regeneration of phase keyed signals ~\cite{Crou2006}.  A theoretical and experimental examination of a single-ended coherent receiver based on a phase sensitive fiber parametric amplifier, including a comparison to a single-ended homodyne detector, has been presented in ~\cite{Kumpera2015}. Our emphasis in this work is on detecting and quantifying unintended phase modulations introduced by commonly used free space optical modulators.”
 
In this paper, we explore the effects of PM on the output of an optical quantum-noise limited PSA, where the phase of the input light is central to amplifier behavior. We study the effects of phase modulation on a PSA both theoretically and experimentally. We introduce a novel method for quantifying the PM depth on an input light field using the PSA as part of a phase-sensitive detector. This method relies on the differing gains of the AC and DC components of the PSA output intensities with a PM input. We compare the results of this method with the results of homodyne detection (HD), a standard method for measuring the quadratures of a light field. We then insert a mechanical chopper in our experiment to amplitude-modulate a laser beam and use our PSA detection method to find PM in this field. We find that this detection method is suitable for detecting phase modulation and that the results for the PSA output match well with our theoretical predictions. We also note that for experiments already employing a PSA, this method allows one to recognize and correct for the presence of PM on the input signal.

\section{Theoretical predictions}
\begin{figure}[t]
    \begin{center}
    \includegraphics[width=\linewidth]{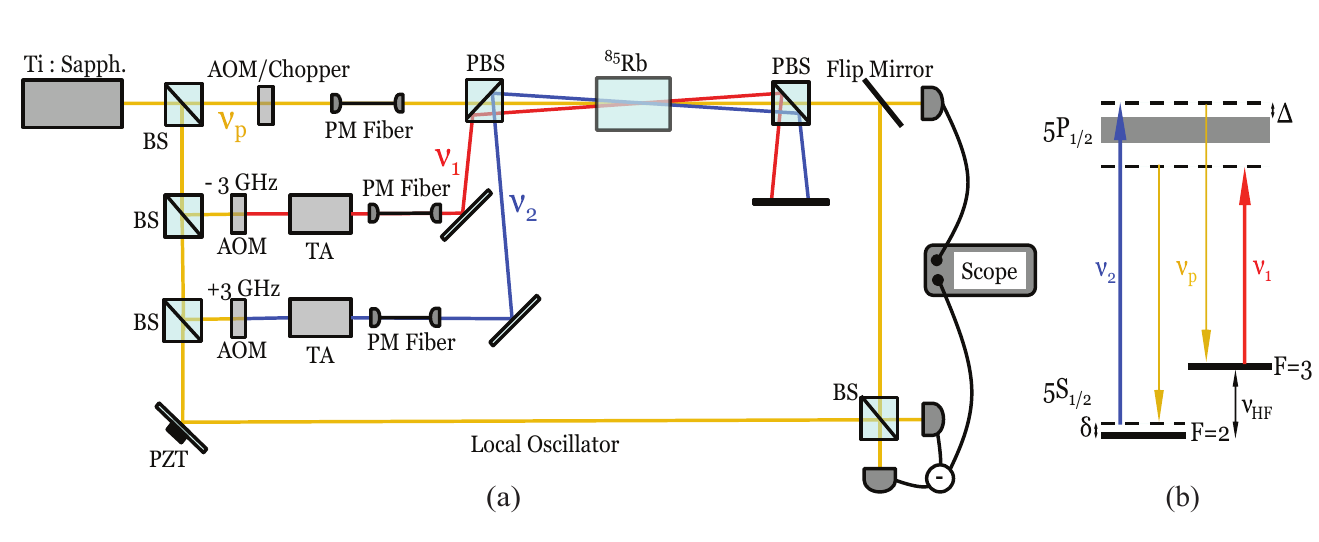}
    \caption{
\label{fig:Experimental_Setup}
    (a) Experimental setup. AOM: acousto-optic modulator, TA: semiconductor tapered amplifier, BS: non-polarizing beam splitter, PBS: polarizing beam splitter. (b) level structure of the D1 transition of $^{85}$Rb and the optical frequencies arranged in the double-$\Lambda$ configuration. Here $\nu_1$ and $\nu_2$ are the pumps and $\nu_p$ is the probe. The width of the excited state in the level diagram represents the Doppler broadened line, $\Delta$ is the one-photon detuning, $\delta$ is the two-photon detuning, and $\nu_{HF}$ is the hyperfine splitting.}
    \end{center}
\end{figure}

\subsection{Phase-sensitive amplification}

We describe the operation of a phase-sensitive amplifier and predict the effect of phase and amplitude modulation on the outputs. Our optical PSA amplifies or deamplifies an optical waveform with a gain dependent on the phase of the input light. The phase of the input field is relative to two strong pump fields (see Fig.~\ref{fig:Experimental_Setup}) that mediate the non-linear process that results in phase-sensitive amplification. Given an input field $E_{in}=|E_{in}|e^{i\phi}$, the relationship between the classical input and output field is given by %
\begin{equation}
 E_{out}= E_{in} \cosh r + E_{in}^* \sinh r , \\
\label{eq: PSA.eq.}
\end{equation} 
where $r$ is the interaction strength of the parametric process derived from the product of the pump power, nonlinear susceptibility, and interaction length \cite{Grynberg:1309862}. In our case, the phase of the process is defined by $\phi_{PSA}=2 \phi-\phi_1-\phi_2$, where $\phi_1$ and $\phi_2$ are the pump phases. Therefore, the output intensity $I_{out}=E_{out}\cdot E_{out}^*$ is a function of input phase $\phi$, making the output phase-sensitive. We define the phase-dependent gain of the PSA as $g(\phi)=I_{out}(\phi)/I_{in}$. This leads to a maximum of the phase-dependent gain of $G = g(0) =  e^{2 r}$ and a minimum of $1/G = g(\pi) = e^{-2 r}$. Because the gain of the PSA changes with input phase, the output intensity will be affected by the presence of phase modulation on the input beam, and this effect must be included.

We now consider the case of a modulated input field. We adopt the definitions used in \cite{bachor_guide_2004} for an electric field in the rotating carrier frame with both amplitude and phase modulation:
\begin{equation}
 E_{in}=[1 - \frac{A}{2}(1-\cos\Omega t) + i\frac{P}{2}\cos\Omega t]e^{i\phi}. \\
\label{eq: Ein_mix.eq.}
\end{equation}
$\Omega$ is the modulation frequency and $A$ and $P$ are the AM and PM modulation depths respectively with $A\ll1$ and $P\ll1$. 
This input field has a sine-wave modulation on top of a constant offset, and so we can refer to the AC and DC components of the field. After the PSA, both the AC and DC components will be amplified or deamplified depending on the input phase $\phi$. However, due to the presence of PM, the AC and DC components can experience different gains. We define $g^{AC}(\phi)=I_{out}^{AC}(\phi)/I_{in}^{AC}$ and $g^{DC}(\phi)=I_{out}^{DC}(\phi)/I_{in}^{DC}$, where $I_{in/out}^{AC}$ and $I_{in/out}^{DC}$ are the AC and DC parts of the input and output intensities, respectively. For pure AM ($P=0$), $g^{AC}(\phi)=g^{DC}(\phi)$, implying that  
the AC and DC components of the input intensity will be equally amplified/deamplified at any given input phase. However, this will not be the case when $P\neq0$, and so we can compare $I_{out}^{DC}(\phi)$ with $I_{out}^{AC}(\phi)$ to detect and quantify phase modulation. 

\begin{figure}[t]
    \begin{center}
    \includegraphics[width=\linewidth]{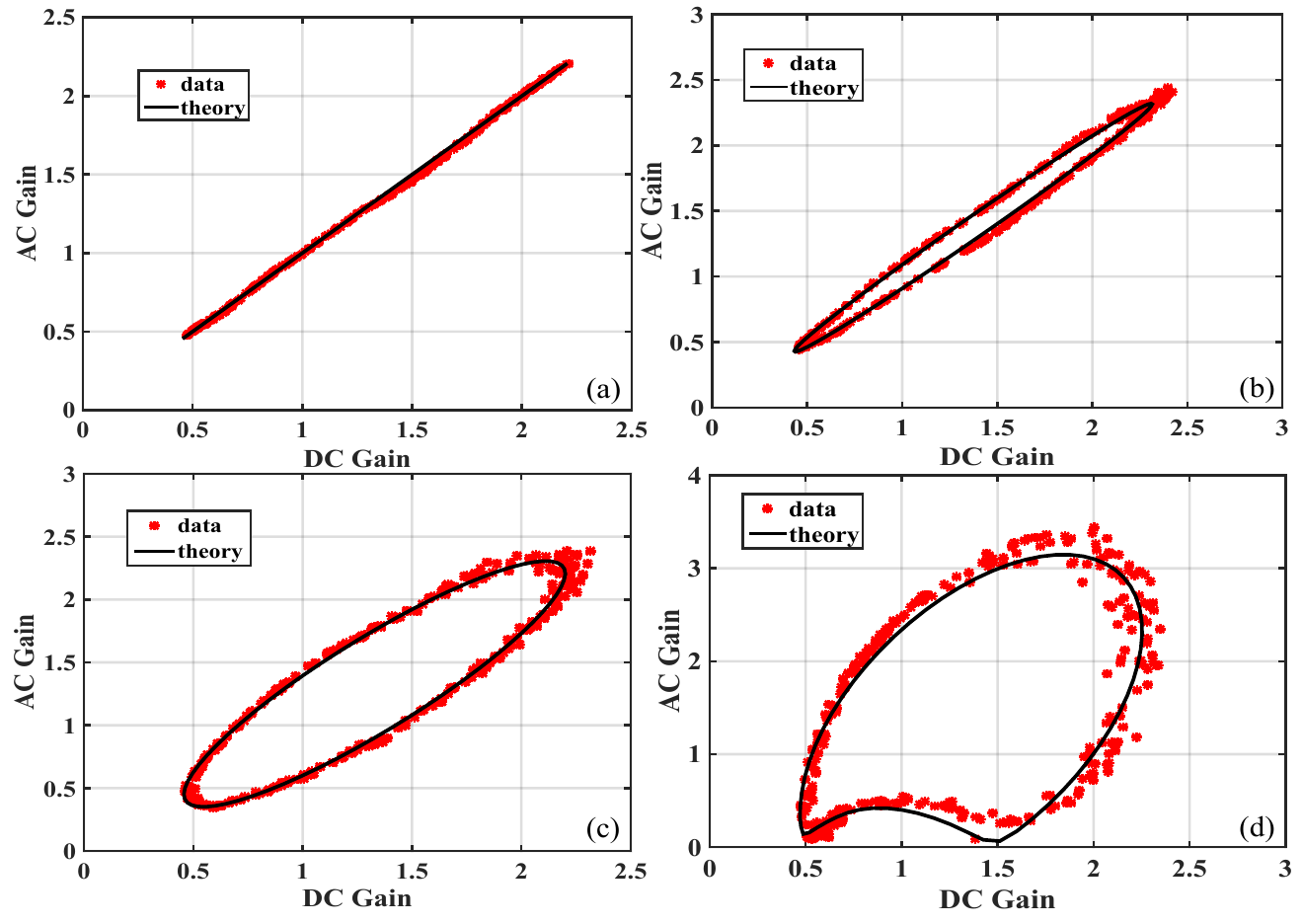}
    \caption{
	\label{fig:acdc}
    PSA results: AC gain versus DC gain for an optical signal modulated with an acousto-optic modulator and amplified in an optical phase-sensitive amplifier. The different plots are for different mixtures of AM and PM due to the AOM alignment. Each plot is parametric with respect to the phase of the PSA. The solid curves are theoretical fits with (a) $P/A=0.00$, (b) $P/A=0.11$, (c) $P/A=0.50$, (d) $P/A=1.65$.}
    \end{center}
\end{figure}

As an example, see the solid lines in Fig. \ref{fig:acdc} which show the PSA AC and DC gains for four different input signals, all with the same level of amplitude modulation but varying levels of phase modulation.  The goal here is to reproduce experimental data (discussed below) from measurements on input signals resulting from four slightly different alignments of an acousto-optic modulator, all producing the same level of AM modulation.  In the experiment, the degree of AM modulation is easily determined by direct intensity measurements without the PSA.  All four theory curves shown here have $A = 0.16$ and a maximum gain of approximately $G=2.25$.


By letting $\phi$ range from 0 to $2\pi$, we get a parametric plot of the AC versus DC gains for all phases. Note that if $P$ is very close to zero (Fig. \ref{fig:acdc}(a)), we see a straight line with a slope of unity when plotting $I_{out}^{DC}$ vs $I_{out}^{AC}$. On the other hand if $P$ is a substantial fraction of $A$ (Figs. \ref{fig:acdc}(b) -- \ref{fig:acdc}(d)) this implies the presence of phase modulation, and we see an oval as the phase is scanned due to the unequal amplifications of the DC and AC components. For large values of $P$ the oval has negative AC gain values but we simply plot the absolute value. Therefore, by detecting the modulated input and output states of a PSA, we can quantify the amount of phase modulation present by using the size and shape of this oval.
\subsection{Balanced homodyne measurement}

We now consider homodyne measurements of a signal with amplitude and phase modulation.  Homodyne detection is the standard phase sensitive technique for measuring the amplitude and phase quadratures of a light field.  We will use it below to look at the same input signals as were measured by the PSA technique (Fig. \ref{fig:acdc}) in order to verify the conclusions drawn from those measurements.”


To perform homodyne measurements, the signal beam is interfered with a reference local oscillator (LO) field $E_{LO}=A_{LO}e^{i\phi_{LO}}$ on a 50/50 beam splitter, after which a balanced detection of the output intensities is performed. Depending on the phase of the reference beam $\phi_{LO}$ compared to the signal phase, the homodyne output will be sensitive to either the amplitude or phase of the signal light. For the same modulated input state in Eq.~(\ref{eq: Ein_mix.eq.}), it can be shown that the subtracted photocurrent is proportional to
\begin{equation}
i^-_{out}= 2A_{LO}(A-2)\sin(\phi_{LO}-\phi) + 2A_{LO}[P\cos(\phi_{LO}-\phi)-A\sin(\phi_{LO}-\phi)]\cos\Omega t. \\
\label{eq: HD.eq.}
\end{equation} 
The first term varies with $\phi_{LO}$ and is independent of the modulation frequency $\Omega$, and will be referred to as the DC level. The second term is dependent on the modulation frequency and will have some AC amplitude.  
When the LO phase satisfies $\phi_{LO}=\phi+n\pi$, where $n$ is an integer, the first term in Eq. (\ref{eq: HD.eq.}) is 0, and the balanced HD output becomes $i^-_{out}= \pm2A_{LO}P\cos\Omega t$. The output is a sine wave at the modulation frequency whose amplitude is determined by only the PM depth $P$ and the LO strength $A_{LO}$. This is the point where the homodyne detector measures the phase quadrature. At $\phi_{LO}=\phi+n\pi/2$, it is sensitive to only the amplitude modulation.  

\begin{figure}[t]
    \begin{center}
    \includegraphics[width=\linewidth]{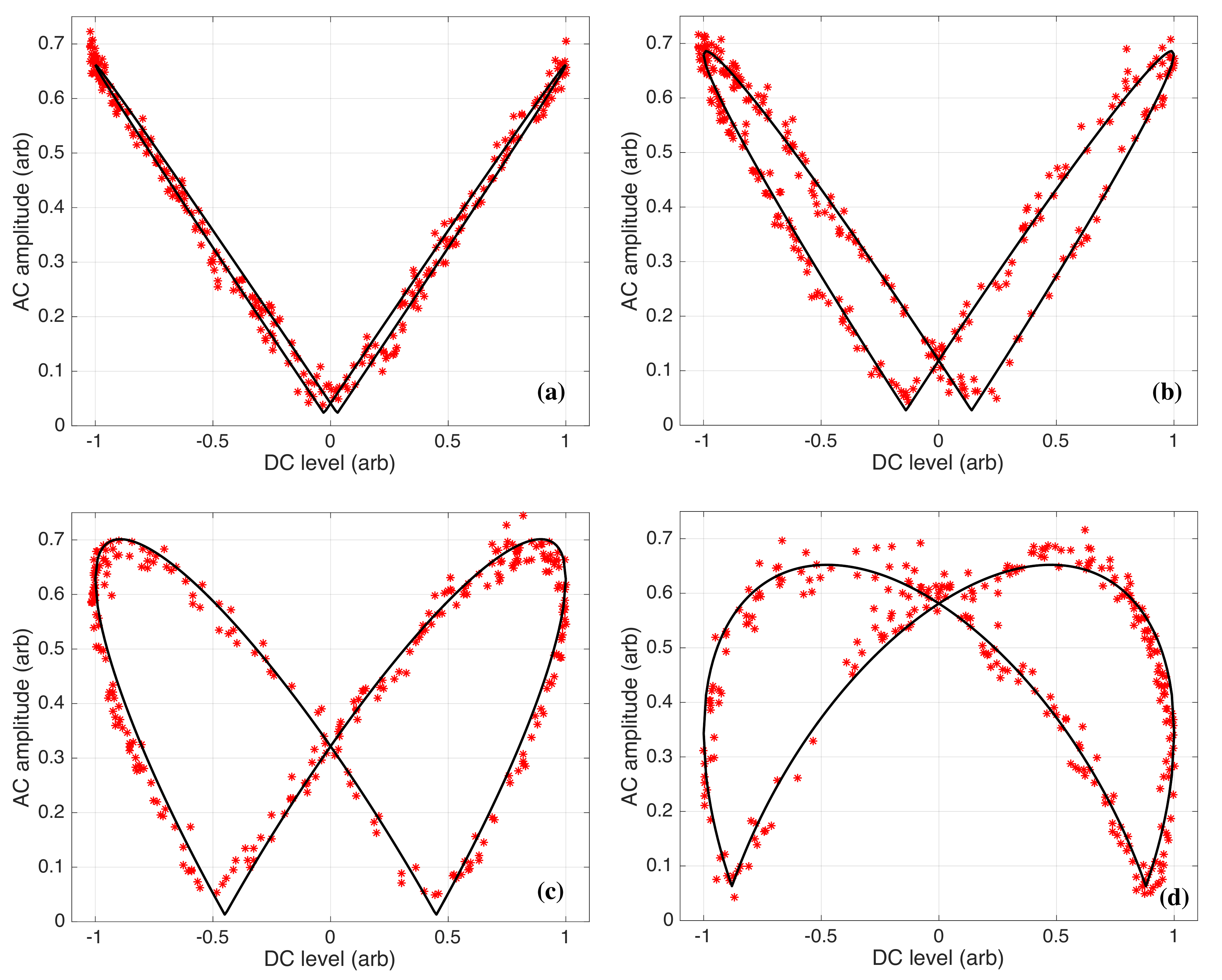}
    \caption{
\label{fig:hd}
   Homodyne results: AC amplitude versus DC level for an optical signal modulated with an acousto-optic modulator and measured with a balanced homodyne detector. The different plots are for different mixtures of AM and PM due to the AOM alignment. Each plot is parametric with respect to the phase of the LO. The solid curves are theoretical fits with (a) $P/A=0.03$, (b) $P/A=0.14$, (c) $P/A=0.50$, (d) $P/A=1.85$.}
    \end{center}
\end{figure}

As an example, in Fig. \ref{fig:hd}, (solid lines) we plot the AC amplitude (second term in Eq. (\ref{eq: HD.eq.})) as a function of the DC level (first term) while $\phi_{LO}$ is scanned over the full range. The data in these plots are for same input signals as in Fig. \ref{fig:acdc}, and thus all have the same level of amplitude modulation ($A=0.16$), and a variable level of phase modulation resulting from slight changes in the alignment of the acousto-optic modulator. The plots are double valued in general because there are two values of $\phi_{LO}$ that correspond to the same DC level.



%

\section{Experiment}
\subsection{Setup}
A diagram of our experiment is shown in Fig.~\ref{fig:Experimental_Setup}. We use an experimental scheme similar to the one detailed in \cite{Corzo:11,CorzoPRL:12}. The phase-sensitive amplifier is created through the nonlinear four-wave mixing process in $^{85}$Rb vapor. The signal probe beam is detuned from the D$_1$ line (795 nm) of Rb while two strong pump beams with frequencies $\pm 3$ GHz from the probe intersect it at a small angle within the atomic vapor. We insert either an 80 MHz AOM or an optical chopper into the probe beam path before the PSA cell to modulate the input light. Before being aligned into the PSA vapor cell, the modulated input beam passes through a single-mode polarization-maintaining fiber. The input probe beam after the fiber is 200 $\mu W$ with a $1/e^2$ beam waist of 250 $\mu m$. The pump beams have a $1/e^2$ beam waist of 550 $\mu m$ and each has a power of 100 $mW$. The 12.5 mm vapor cell is filled with isotopically pure $^{85}$Rb and heated to 87 $^\circ$C. All the data shown in this paper is taken with the probe beam blue detuned 1.4 GHz from the center of the  $5S_{1/2}$ $F=3$ manifold to the center of the $5P_{1/2}$ Doppler-broadened transition. The pump beams are created by seeding two 0.5 $W$ tapered amplifiers with light that has been shifted $\pm$3 GHz using double-passed AOMs. The probe frequency is always centered between the two pumps.  The chosen detunings result in a $-4$ MHz two-photon detuning for the probe and each pump compared to the exact hyperfine splitting of the ground state, in order to compensate for light shifts (see Fig.~\ref{fig:Experimental_Setup}(b)). 
\subsection{Results: AOM}

We have found that when using the AOM to amplitude-modulate a light beam, the amount of  (unintended) PM is highly dependent on the AOM alignment relative to the input beam while the degree of amplitude modulation is not. For present purposes the degree of amplitude modulation can be determined by direct intensity detection without the PSA or homodyne detector and thus we take it as a fixed parameter in our fits to the PSA and homodyne data.
For a given alignment through the AOM, we can switch between detecting the light using HD or sending it into the PSA. The input beam is modulated at $1$ MHz with $A=0.16$.

Figures \ref{fig:acdc} and \ref{fig:hd} show measurements of PM at four different AOM alignments using the PSA method and HD method respectively. The alignment is changed by moving the horizontal tilt on the AOM. The stars are the experimental data while the solid lines are the theoretical fits using Eqs.~(\ref{eq: PSA.eq.}), (\ref{eq: Ein_mix.eq.}) and (\ref{eq: HD.eq.}). 

\begin{figure}[H]
    \begin{center}
    \includegraphics[width=4in]{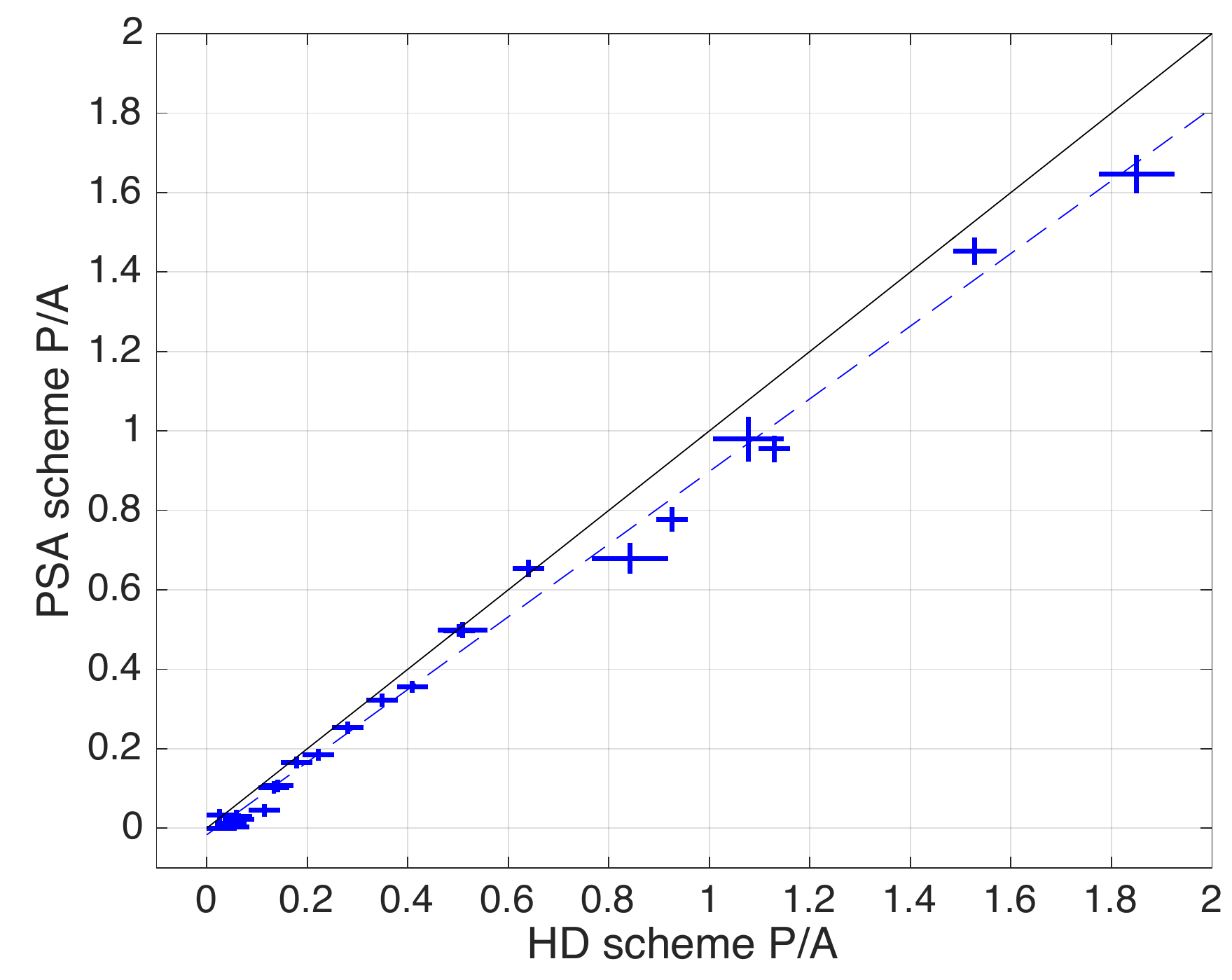}
    \caption{
\label{fig:ACDC_HD_Comparison}
		Comparison of the ratio of phase to amplitude modulation measured by homodyne detection and measurements of PSA AC and DC amplification. The solid line is $y=x$ and the dashed line is a best linear fit, $y = 0.91x  - 0.02$,  to the data.}
    \end{center}
\end{figure}

For the data in Fig. \ref{fig:acdc}, the input and pump phases are allowed to drift such that each data point represents a shot of the experiment at a different PSA phase, and therefore a different $g(\phi)$. The PSA measurements give a well-defined shape which is a function of $A$, $P$, $\phi$, and $r$. 
To fit the data, we took a subset of the data that could be plotted using a single-valued function of AC gain as a function of DC gain, rather than a parametric function of $\phi$. To get a single valued function, we selected the data points corresponding to a span of $\pi$ in PSA phase, which can be found by taking all the data points that lie above the line $y=x$ in Fig. \ref{fig:acdc}. This data is then fit using the AC gain as a function of DC gain. The uncertainties for these measurements can be found in Fig. \ref{fig:ACDC_HD_Comparison}. The uncertainties are $95\%$ confidence intervals from the fits. Due to systematic errors and our initial uncertainty in $A$, we have put lower bounds on the uncertainties corresponding to $1.5\%$  uncertainty in $P/A$.

We find that by moving the tilt of the AOM less than one degree, we can change $P/A$ from nearly zero to greater than 0.2. The amplitude modulation alone is not appreciably changed for any of the data shown. Extremely fine tuning of the angle is required to find the minimum $P/A$. Unfortunately, aligning for highest diffraction efficiency does not guarantee minimum phase modulation.

In the case of the homodyne measurements, the shapes are a function of $P$ and $A$ from the second term of Eq.~(\ref{eq: HD.eq.}), as well as a scaling factor and a vertical offset.
To perform fits, we took a subset of the data that could be plotted using a single-valued function. To select data points consistent with a single valued function, we selected the data points corresponding to a span of $\pi$ in LO phase. Moving along the parametric curve in a single direction, from the minimum DC level to the maximum DC level, constitutes a $\pi$ phase shift in the LO. Instead of a parametric function of $\phi$, we can now plot AC amplitude as a function of DC level and perform a standard fit. The uncertainties for these measurements can be found in Fig. \ref{fig:ACDC_HD_Comparison}. The uncertainties are $95\%$ confidence intervals from the fits. Due to systematic errors and our initial uncertainty of $A$, we have put lower bounds on the uncertainties corresponding to $3\%$  uncertainty in $P/A$.

By switching between the PSA measurement and the homodyne measurement without disturbing the AOM alignment, we can compare the two methods. In Fig. \ref{fig:ACDC_HD_Comparison}, we plot $P/A$ for the homodyne measurement versus the PSA measurement. The AOM alignment was adjusted for each point to increase or decrease the amount of phase modulation, and thus the ratio $P/A$. We find that the PSA and homodyne method track each other linearly and are in substantial agreement. This shows that measurements with a phase-sensitive amplifier can act as a diagnostic tool for reducing phase modulation on a light field.

%
\subsection{Results: optical chopper}
Having confirmed that our analysis of the PSA data is consistent with that based on standard homodyne measurements, we can now use the PSA results alone.  We consider here a common laboratory technique for amplitude modulating a beam, namely a mechanical chopper wheel which alternately blocks and un-blocks a beam.  The spatial mode of the beam after the chopper is cleaned with a single mode fiber before the beam is sent into the PSA.   
We find the chopper introduces phase modulation as its blades cut through the beam. It may seem counter-intuitive that a chopper can add PM; however, as the blade passes through the beam, the spatial mode and phase front of the beam are disturbed due to diffraction effects around the blade. As the chopper moves through the beam, the light intensity will change with a transient, well-modeled by the error function. 

\begin{figure}[H]
    \begin{center}
    \includegraphics[width=\linewidth]{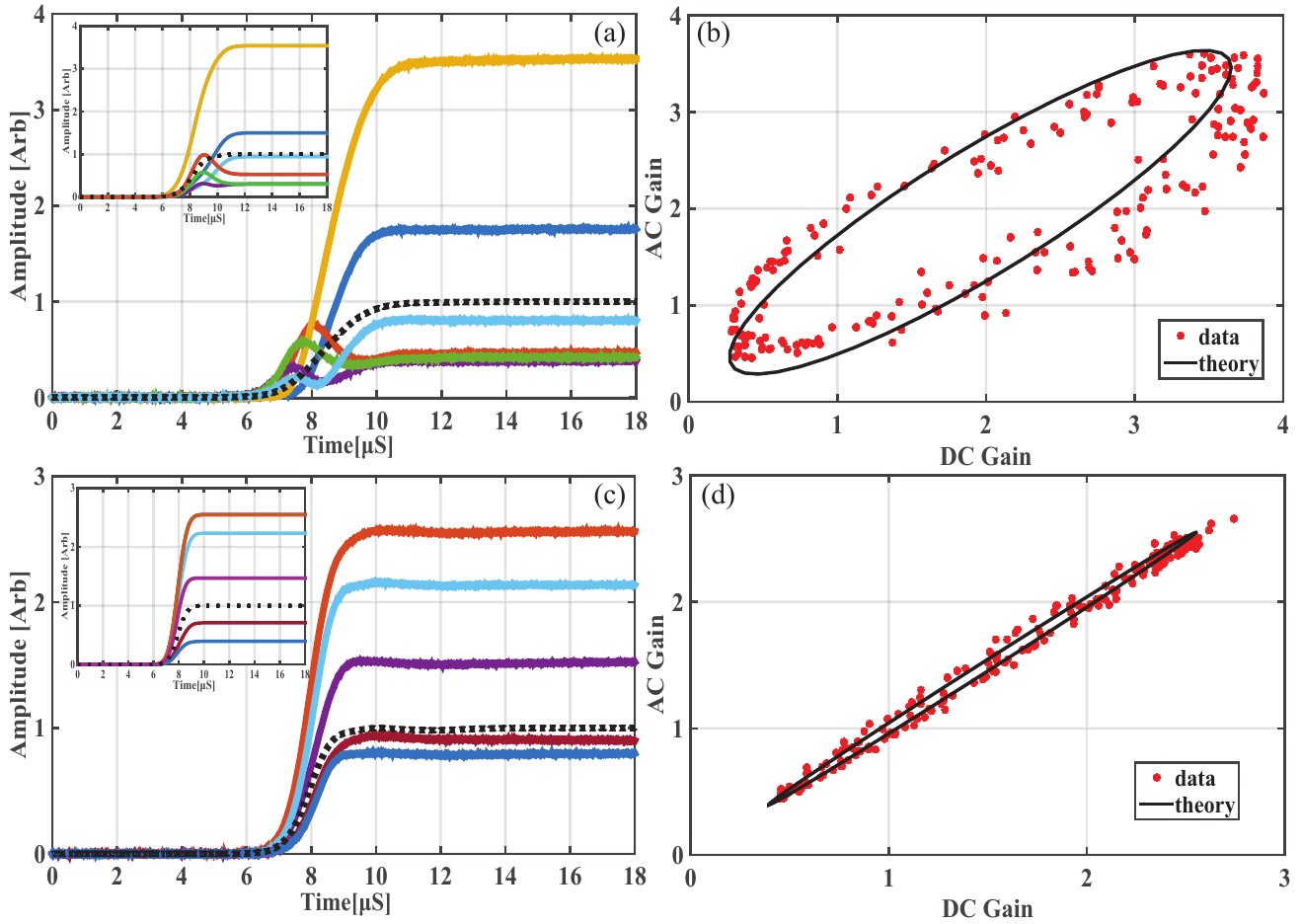}
    \caption{
\label{fig:chopper}
		PM measurements for two different chopper alignments using the PSA scheme. (a) and (c) show raw data from a tilted chopper alignment and optimal chopper alignment, respectively. The dashed lines are direct intensity detection without a PSA and the other curves are various phases of the PSA. Inset theory curves are shown as examples to demonstrate curve shapes for the fit parameters and do not necessarily match the PSA phases of the individual data curves shown. (b) and (d) show AC gain vs. DC gain, as defined in the text, for the tilted chopper alignment, and optimal chopper alignment, respectively. The solid curves in (b) and (d) are theoretical fits where $P$ = 0.7 and $P$ = 0.15, respectively.}
    \end{center}
\end{figure}

The intensity as a function of time after a blade moving through a Gaussian beam is given by
\begin{equation}
 I_{in} = \frac{1}{2}[1+erf(\frac{t-\mu}{\sqrt{2}\sigma})], \\
\label{eq: erf.}
\end{equation} 
where $t$ is time, $\mu$ is the offset and $\sigma$ is the width of the error function. We introduce a simple empirical model for the phase modulation which we can test with the PSA measurements. We assume that any phase modulation introduced to the light follows a Gaussian function in time as the blade traverses across the beam profile, with a width that matches the error function of the intensity: 
\begin{equation}
 E_{in}= e^{i\phi}\sqrt{\frac{1}{2}[1+iPe^{-(\frac{t-\mu}{\sqrt{2}\sigma})^2}][1+erf(\frac{t-\mu}{\sqrt{2}\sigma})]}. \\
\label{eq: chopper.eq.}
\end{equation} 
In this case, $P$ corresponds to the maximum amplitude of the Gaussian shaped PM. We find the values of $\mu$ and $\sigma$ by fitting Eq.~(\ref{eq: erf.}) to the measured transient intensity of the input beam (see the dashed curves in Figs. \ref{fig:chopper}(a) and \ref{fig:chopper}(c)). We define the AC part of the PSA output signal as the intensity integrated over the time window during which the transient turns from off to on; from 6 to 10 $\mu s$. The AC gain is defined as the ratio of the AC signal with the PSA on to the AC signal with the PSA off. The DC component is the steady state of the intensity after the light is fully unblocked.The DC gain is defined as the ratio of the DC signal with the PSA on to the DC signal with the PSA off. Just as when using an AOM, the discrepancy between AC and DC gain is indicative of the level of phase modulation. This allows us to plot the AC intensity component versus the DC level and extract a PM depth. We believe the deviation between theory and experiment is mostly due to our assumption of the Gaussian form of the phase modulation above and the slight mechanical instability of the chopper from shot-to-shot. 

We found that the phase modulation of a chopper depended strongly on the tilt of the chopper blades when they intersected the laser beam. 
Fig.~\ref{fig:chopper} shows the results of the PSA measurements for two positions of an optical chopper, the first (\ref{fig:chopper}(a) and \ref{fig:chopper}(b)) when the blades are tilted off-axis by approximately $10$ degrees from the beam path, and the second (\ref{fig:chopper}(c) and \ref{fig:chopper}(d)) where the blades intersect the beam path at normal incidence. In both cases, the blades are placed within the Rayleigh range of a beam focus. It is clear from this measurement that the phase modulation was reduced by setting the chopper to normal incidence, however PM may not be eliminated completely. We were unable to reduce the amount of PM below the level shown in Fig.~\ref{fig:chopper}(d).

\section{Conclusion}


These demonstrations highlight the importance of being able to measure and correct for the presence of unintended phase modulation when employing common amplitude modulation techniques in experiments using phase sensitive amplifiers.  We show that PSA signals can be used as a diagnostic tool for quantifying the PM depth of an input signal and are consistent with established homodyne techniques.  
We find that both AOMs and optical choppers can inadvertently add PM to a light field in addition to the desired AM. This can drastically alter the results in applications using phase-sensitive amplifiers. In each case, the amount of PM can be reduced by adjusting the angle of incidence between the beam path and the modulator. Similar analysis could be carried out using optical PSAs and light modulated by electro-optic devices.

\section*{Funding}
National Science Foundation (NSF) and Air Force Office of Scientific Research (AFOSR).

%


\begin{thebibliography}{99}

\bibitem{PhotonicDevices} J.-M. Liu, {\it Photonic Devices} (Cambridge University, 2005).

\bibitem{Whittaker:85} E. A. Whittaker, M. Gehrtz, and G. C. Bjorklund, ``Residual amplitude modulation in laser electro-optic phase modulation'' J. Opt. Soc. Am. B, {\bf 2}(8), 1320-1326 (1985).

\bibitem{Cusack:04} B. J. Cusack, B. S. Sheard, D. A. Shaddock, M. B. Gray, P. K. Lam, and S. E. Whitcomb, ``Electro-optic modulator capable of generating simultaneous amplitude and phase modulations'' Appl. Opt., {\bf 43}(26), 5079-5091 (2004).

\bibitem{HENDERSON1970223} D. M. Henderson and R. L. Abrams, ``A comparison of acoustooptic and electrooptic modulators at 10.6 microns'' Opt. Commun., {\bf 2}(5), 223-226 (1970).

\bibitem{Li_OptLett2005} E. Li, J. Yao, D. Yu, J. Xi, and J. Chicharo, ``Optical phase shifting with acousto-optic devices'' Opt. Lett., {\bf 30}(2), 189-191 (2005).

\bibitem{Corzo:11} N. Corzo, A. M. Marino, K. M. Jones, and P. D. Lett, ``Multi-spatial-mode single-beam quadrature squeezed states of light from four-wave mixing in hot rubidium vapor'' Opt. Express, {\bf 19}(22), 21358-21369 (2011).

\bibitem{CorzoPRL:12} N. V. Corzo, A. M. Marino, K. M. Jones, and P. D. Lett, ``Noiseless optical amplifier operating on hundreds of spatial modes'' Phys. Rev. Lett., {\bf 109}, 043602 (2012).

\bibitem{bachor_guide_2004} H. A. Bachor and T. C. Ralph, {\it A Guide to Experiments in Quantum Optics, Second Revised and Enlarged Edition} (Wiley-VCH, 2004).

\bibitem{Yam:15} W. Yam, E. Davis, S. Ackley, M. Evans, and N. Mavalvala, ``Continuously tunable modulation scheme for precision control of optical cavities with variable detuning'' Opt. Lett., {\bf 40}(15), 3675-3678 (2015).

\bibitem{PICHE1988146} M. Piché, C. Par, and P.-A. Blanger, ``Conversion of phase modulation to amplitude modulation using a phase conjugate mirror'' Opt. Commun., {\bf 65}(2), 146-150 (1988).

\bibitem{Yao98} X. S. Yao, ``Phase-to-amplitude modulation conversion using Brillouin selective sideband amplification'' IEEE Photon. Tech. Lett., {\bf 10}(2), 264-266 (1998).

\bibitem{Crou2006} K. Croussore, I. Kim, C. Kim, Y. Han, G. F. Li, ``Phase-and-amplitude regeneration of differential phase-shift keyed signals using a phase-sensitive amplifier,'' Opt. Express {\bf 14}(6), 2085-2094 (2006).

\bibitem{Kumpera2015} A. Kumpera, R. Malik, A. Lorences-Riesgo, and P.A. Andrekson, ``Parametric coherent receiver,'' Opt. Express {\bf 23}(10), 12952-12964 (2015).

\bibitem{Grynberg:1309862} G. Grynberg, A. Aspect, and C. Fabre, {\it Introduction to Quantum Optics: From the Semi-classical Approach to Quantized Light} (Cambridge University, 2010).

\end{thebibliography}
\end{document}